\documentclass[12pt]{article}
\pdfoutput=1
\usepackage{graphicx}
\usepackage{color}
\usepackage{cancel}
\usepackage{comment}
\usepackage{amsmath,amssymb,booktabs,xcolor,cite}

\newcommand{\beq}{\begin{eqnarray}} 
\newcommand{\eeq}{\end{eqnarray}}

\begin{document}

\begin{flushright}
{\tt KCL-PH-TH/2026-30, CERN-TH-2026-306}
\end{flushright} 

\vspace*{5mm} 

\begin{center}

\mbox{\large\bf  A pseudoscalar Higgs boson interfering with toponium:}

\vspace*{2mm}

\mbox{\large\bf   a QCD analysis of $\mathbf{gg\to A\to t\bar t}$ near threshold} 

\vspace*{8mm}

{\sc Abdelhak~Djouadi$^{1                 }$} and  {\sc John Ellis$^{2,3}$} 

\vspace{6mm}

{\small 
\mbox{ \hspace*{-6mm} $^1$ Departamento de F\'isica Te\'orica y del Cosmos, Universidad de Granada,
18071 Granada, Spain.} \\[2mm] 

$^2$ Theoretical Physics Department, CERN, CH 1211 Geneva, Switzerland.\\[2mm]

$^3$ Physics Department, King’s College London, Strand, London, WC2R 2LS, United Kingdom.
}

\end{center}

\vspace*{8mm}

\begin{abstract}
\noindent
The CMS and ATLAS collaborations have observed an excess of $t\bar t$ events near threshold that is consistent with the formation of the pseudoscalar toponium state $\eta_t$. A pseudoscalar Higgs boson $A$ with a large coupling to top quarks could contribute part of this excess and the presence of the $A$ boson could be concealed. We point out that, if such a state exists with $M_A$ close to $2m_t$, it necessarily interferes with toponium, since both live in the colour--singlet ${}^1S_0$ channel and share the same Coulomb final--state interaction. Using exact leading--order amplitudes, whose relative phase is fixed by unitarity, and Coulomb resummation applied coherently to the QCD background, the $A$ boson signal and their interference, we show that the line shape is controlled by a single complex number $\kappa$, the ratio of the resonant and non--resonant amplitudes at the peak. Unitarity fixes ${\rm Im}\,\kappa=-1$ exactly when $A$ decays into top quarks, so that the interference reduces the resonant rate by $2/|\kappa|^2$ and produces an exact zero of the ${}^1S_0$ amplitude just above $M_A$. The parameter $\kappa$ does not depend on the $A$ coupling to top quarks and, with the Coulomb--enhanced width we find $|\kappa|\approx1.5$--$1.7$ for $M_A=350$--$380$ GeV and, hence, the $A$ resonance is completely distorted. For an $A$ boson accounting for 50\% (25\%) of the observed excess, it enhances the toponium peak by about 10 to 70\% (5 to 35\%), and the threshold excess differs strongly from the sum of the toponium and $A$ contributions.
\end{abstract}

\newpage

\subsection*{1. Introduction}

The CMS collaboration has observed an excess of $t\bar t$ events near the production threshold in the dilepton channel, with the spin correlations of a pseudoscalar state~\cite{CMS:HIG22013,CMS:toponium}. The excess has since been confirmed by CMS in the single--lepton channel, with a significance of 6.1 standard deviations and a cross section of $5.1\pm0.9$ pb in a simplified 
model~\cite{CMS:ljets}, and its existence has been confirmed by the ATLAS collaboration~\cite{ATLAS:toponium}. Its most natural interpretation is the colour--singlet ${}^1S_0$ quasi--bound state of toponium, $\eta_t$, predicted by QCD~\cite{Fadin:1987wz,Kiyo:2008bv,Sumino:2010bv,Fuks:2021xje}. 
An alternative interpretation is a beyond the Standard Model (SM) pseudoscalar boson $A$ which has been searched for in the same $t\bar t$ final state~\cite{CMS:AHtt}. Such a possibility has been considered in Ref.~\cite{DEQ:2024} using an effective model in which the $A$ coupling to top quarks is 
\beq
{\cal L}\supset i\,g_{Att}\, ({m_t}/{v})\,\bar t\gamma_5t\,A\, ,
\eeq
A major issue was that the main $A$ production channel at the LHC is gluon fusion, $gg \to A$ 
\cite{Spira:1995rr}, and for masses near or above the $t\bar t$ kinematical  threshold, the main decay mode  is by far $A\to t\bar t$; the $A$ signal will thus interfere with the continuum QCD background,  $gg \to t\bar t$ \cite{Dicus:1994bm,DEQ:2016,Maltoni:2024}. In Ref.~\cite{DEQ:2024}, we have thus studied the impact of the interference, simply working in perturbative QCD at fixed order, as in the studies on the subject that we performed  earlier. In that analysis, the toponium was not included, and the couplings $g_{Att}$ were chosen such that the $A$ boson alone accounts for all of the excess observed by CMS. 

Now that the excess has been confirmed in several channels and is well described by toponium, an $A$ boson accounting for the whole excess is clearly  disfavored: if $A$ exists, it can only account for a fraction of it, the rest being due to toponium. But if both states exist, they interfere. The $A$ boson couples to a $t\bar t$ pair in a colour singlet with $J^{\rm PC}=0^{-+}$, i.e., in the ${}^1S_0$ state, which is exactly the state that forms $\eta_t$. Near threshold, the QCD amplitude in this channel is dominated by the exchange of Coulomb gluons between the slowly-moving top quarks; its resummation produces the toponium enhancement, but it applies equally to the $A$ final state decay. The interference of $A$ with the QCD background therefore \emph{is} its interference with toponium, once the Coulomb interaction is included. 

In this Letter we illustrate the importance of this interference for $A$ bosons accounting for 50\% and 25\% of the observed excess, keeping the case of 100\% as a reference, in the same framework as our recent study of the interference effects in the heavy MSSM Higgs searches in the $gg\to H\to WW,ZZ,hh$ and $gg\to A\to Zh$ channels~\cite{DE:2026}. In that paper, we have seen that the line shape is described by a Fano profile~\cite{Fano} controlled by a single complex number $\kappa$. We adopt this interpretation here and we compute the parameter $\kappa$ for the the joint toponium-$A$  
system at fixed order and with Coulomb resummation.

It is shown that the effects of this interference between toponium and the $A$ boson may be quite substantial, potentially enhancing the toponium signal and distorting any accompanying $A$ signal. The analysis we present here provides a formalism that may be useful for future experimental studies of the $t \bar t$ threshold to clarify the nature of the observed enhancement and constrain the possibility of a nearby $A$ boson. {It could also be useful  in future searches near the $t\bar t $ threshold of any pseudoscalar Higgs boson predicted by theories beyond the SM like the MSSM or a two-Higgs doublet model.}

{The paper is organised as follows: in the next section we summarize the parameterisation used for the interference and  discuss the Coulomb resummation for the toponium. In section 3, we present our numerical results and, in section 4,  our conclusions.}

\subsection*{2. Interference in the $\kappa$ parametrization and toponium}

{As in our recent study of interference effects in the MSSM Higgs sector~\cite{DE:2026} and following a formalism introduced long ago by Fano in Ref.~\cite{Fano},} we will write the amplitude for $gg\to t\bar t$ in the channel to which the resonance couples,  as a pole term with residue ${\cal R}$ plus a smooth non--resonant part ${\cal N}(\hat s)$, where the QCD amplitude is projected on the colour--singlet ${}^1S_0$ state. The line shape relative to the non--resonant background is then
\beq
\frac{|{\cal A}|^2}{|{\cal N}|^2}=\left|1+\frac{\kappa}{\epsilon+i}\right|^2,\qquad \epsilon\equiv\frac{\hat s-M_A^2}{M_A\Gamma_A}\, ,\qquad \kappa\equiv\frac{\cal R}{M_A\Gamma_A\,{\cal N}(M_A^2)}\, .
\label{eq:kappa}
\eeq
The pure resonance, added without interference, has height $|\kappa|^2$ relative to the background. The real part of $\kappa$ produces an antisymmetric peak--dip structure, which integrates to almost zero, while its imaginary part produces a symmetric term with the shape of the signal. Integrating over $\epsilon$, the signal gives $\pi|\kappa|^2$ and the interference $2\pi\,{\rm Im}\,\kappa$, so that the interference changes the resonant rate by $\delta=2\,{\rm Im}\,\kappa/|\kappa|^2$ relative to the signal.

We have calculated the helicity amplitudes for the signal process $gg\to A\to t\bar t$ and for the QCD background $gg\to t\bar t$  using the exact top-quark loop with explicit colour and spin sums; the angle--integrated QCD cross section agrees with the analytic LO formula and its colour--singlet fraction tends to $2/7$ at threshold as it should. The relative phase of the two amplitudes, which determines the sign of the interference, is fixed by unitarity: the absorptive part of the $gg\to A$ loop must equal the $t\bar t$ cut of the tree--level QCD amplitude times the $A\to t\bar t$ vertex, a relation our amplitudes satisfy to high accuracy.

A direct consequence is that when $A$ decays only into top quarks, one has
\beq
{\rm Im}\,\kappa=-1\, ,\qquad \delta=-{2}/{|\kappa|^2}\, ,
\label{eq:Imkappa}
\eeq
independently of the coupling $g_{Att}$ and of $M_A$. {To understand why ${\rm Im}\,\kappa=-1$},  consider the ${}^1S_0$ colour--singlet partial wave, in which the $A$ boson couples to both the $gg$ initial state and the $t\bar t$ final state. The resonant amplitude is 
given by
\beq
{\cal A}_{\rm res}=g_{Agg}g_{Att}/(\hat s-M_A^2+iM_A\Gamma_A) , 
\eeq 
where the effective $Agg$ coupling $g_{Agg}$ is generated by the top loop. Its absorptive part comes from the on--shell $t\bar t$ intermediate state, ${\rm Im}\,g_{Agg}={\cal N}\,g_{Att}\,\Phi_{t\bar t}$, where ${\cal N}$ is the non--resonant $gg\to t\bar t$ amplitude in this partial wave and $\Phi_{t\bar t}$ the two--body phase space, while the width is $M_A\Gamma_A=g_{Att}^2\Phi_{t\bar t}$ when $A$ decays only into $t\bar t$ final states. Writing the $Agg$ coupling as $g_{Agg}={\rm Re}\,g_{Agg}+i\,{\rm Im}\,g_{Agg}$, one obtains
\beq
\kappa=\frac{g_{Agg}\,g_{Att}}{M_A\Gamma_A\,{\cal N}}=\frac{{\rm Re}\,g_{Agg}\,g_{Att}}{M_A\Gamma_A\,{\cal N}}+i\,\frac{{\rm Im}\,g_{Agg}\,g_{Att}}{g_{Att}^2\,\Phi_{t\bar t}\,{\cal N}}\, ,\qquad \left|{\rm Im}\,\kappa\right|=\frac{|{\cal N}|\,g_{Att}^2\,\Phi_{t\bar t}}{g_{Att}^2\,\Phi_{t\bar t}\,|{\cal N}|}=1\, .
\eeq
The sign is fixed by unitarity as well: the total rate in the ${}^1S_0$ channel cannot exceed the unitarity bound, which requires the interference to be destructive on average, $\delta<0$, i.e., ${\rm Im}\,\kappa=-1$ in the convention of eq.~(\ref{eq:kappa}), as found numerically in Table~\ref{tab:kappa}. The real part of $\kappa$ comes from the dispersive part of the top loop, and is not fixed by unitarity. The relation is unchanged by the Coulomb resummation, which multiplies the absorptive part and the width by the same factor $K_1/\beta$. It is the analogue, for the production of a resonance in a two--channel problem dominated by one decay mode, of the elastic unitarity relation which, for a pure Breit--Wigner resonance in elastic scattering, fixes the phase of the resonant amplitude relative to the background and leads to the Fano zero~\cite{Fano}.

The rate--changing part of the interference is thus completely fixed by unitarity, and the model dependence resides entirely in the term ${\rm Re}\,\kappa$, which comes from the dispersive part of the top loop. A further consequence is that the ${}^1S_0$ amplitude in eq.~(\ref{eq:kappa}) vanishes exactly at the value $\epsilon=-{\rm Re}\,\kappa$: this is the classic Fano zero. This situation is very different from the $gg\to H\to VV$ channels with $V=W,Z$ studied in Ref.~\cite{DE:2026}, where ${\rm BR}(H\to VV)\ll1$ and the interference is a modest correction.


Since both the residue ${\cal R}$ and the width $\Gamma_A$ are proportional to $g_{Att}^2$ when $A$ decays only into top quarks, $\kappa$ is independent of $g_{Att}$. The profile of eq.~(\ref{eq:kappa}), as a function of $\epsilon$, is therefore the same for all couplings: reducing the coupling only makes the resonance narrower, its integrated signal scaling like $g_{Att}^2$. The values of $\kappa$ for the three $A$ masses that we use as benchmarks in our study, namely $M_A\!=\!350, 365$ and 380 GeV, i.e. below and above the threshold region, are given in Table~\ref{tab:kappa}. At fixed order, $|\kappa|$ ranges from 4.7 for $M_A=350$ GeV, where the width is small because of the $\beta_t$ suppression, to 1.9 for $M_A=380$ GeV; the interference reduces the narrow--width rate by 9 to 55\%.

For each of our three benchmarks for the $A$ mass, $M_A\!=\!350, 365$ and 380 GeV, we define the coupling $g_{Att}^{100\%}$ for which the $A$ signal alone accounts for the whole excess observed by CMS in the dilepton channel, as in our previous analysis~\cite{DEQ:2024}, and the couplings for which it accounts for a fraction $f$ of it: $g_{Att}=\sqrt f\,g_{Att}^{100\%}$ with the illustrative choices $f=0.5$ and 0.25, yielding an integrated signal scaling like $g_{Att}^2$. The benchmarks are given in Table~\ref{tab:bench}; the case $f=50\%$ will be used for illustration in the figures.

\begin{table}[!h]
\renewcommand{\arraystretch}{1.2}
\centering\small\setlength{\tabcolsep}{6pt}
\begin{tabular}{c ccc ccc}
\toprule
 & \multicolumn{3}{c}{$g_{Att}$} & \multicolumn{3}{c}{$\Gamma_A\,$[GeV], fixed$\,$order/Coulomb resummed}\\
$M_A$ [GeV] & $f=100\%$ & $50\%$ & $25\%$ &  \hspace*{7mm} $f=100\%$ & $50\%$ & $25\%$\\
\midrule
350 & 0.81 & 0.57 & 0.41 & \hspace*{9mm} 2.3 / 7.9 & 1.1 / 3.9 & 0.6 / 2.0\\
365 & 0.78 & 0.55 & 0.39 & \hspace*{9mm} 4.3 / 8.2 & 2.1 / 4.1 & 1.1 / 2.1\\
380 & 0.83 & 0.59 & 0.42 & \hspace*{9mm} 6.4 / 7.6 & 3.2 / 3.8 & 1.6 / 1.9\\
\bottomrule
\end{tabular}
\caption{The benchmark masses $M_A=350, 365$ and 380 GeV and the couplings $g_{Att}$ for which the $A$ boson accounts for a fraction $f=100\%$, 50\% and 25\% of the threshold excess, and the corresponding total widths at fixed order and with the Coulomb resummation.}
\label{tab:bench}
\end{table}

\begin{table}[!h]
\renewcommand{\arraystretch}{1.2}
\centering\small\setlength{\tabcolsep}{5pt}
\begin{tabular}{l c ccc c ccc}
\toprule
 & & \multicolumn{3}{c}{fixed order} & & \multicolumn{3}{c}{Coulomb resummed}\\
$M_A$ [GeV] & & $\Gamma_A^{100\%}$ [GeV] & $\kappa$ & $\delta$ & & $\Gamma_A^{100\%}$ [GeV] & $\kappa$ & $\delta$\\
\midrule
350 & & 2.3 & $-4.56-1.00\,i$ & $-0.09$ & & 7.9 & $-1.32-1.00\,i$ & $-0.73$\\
365 & & 4.3 & $-2.21-1.00\,i$ & $-0.34$ & & 8.2 & $-1.14-1.00\,i$ & $-0.87$\\
380 & & 6.4 & $-1.62-1.00\,i$ & $-0.55$ & & 7.6 & $-1.37-1.00\,i$ & $-0.70$\\
\bottomrule
\end{tabular}
\caption{The total width, the interference parameter $\kappa$ of eq.~(\ref{eq:kappa}) and the narrow--width rate change $\delta=2\,{\rm Im}\,\kappa/|\kappa|^2$ for the three $A$ masses (independently of $g_{Att}$), at fixed order and with the Coulomb resummation and unitarity--consistent width.}
\label{tab:kappa}
\end{table}

{Let us now discuss the Coulomb resummation and toponium.} 
Near threshold, both the QCD and the $A$ amplitudes produce the $t\bar t$ pair at short distances, $\sim1/m_t$, after which the pair propagates non--relativistically in the Coulomb potential $V_R=-C_R\alpha_s/r$, with $C_1=\frac43$ for the colour singlet and $C_8=-\frac16$ for the octet. The cross section factorizes as
\beq
\frac{d\sigma_R}{dm_{t\bar t}}=\left.\frac{d\sigma_R^{\rm Born}}{dm_{t\bar t}}\right|_{\beta\to K_R(E)}\, ,\qquad K_R(E)=\frac{4\pi}{m_t^2}\,{\rm Im}\,G_R(E+i\Gamma_t)\, ,\qquad E=m_{t\bar t}-2m_t\, ,
\eeq
where $G_R$ is the Coulomb Green function at the origin, which we evaluate as the sum over bound states plus the Sommerfeld--enhanced continuum, smeared by the top quark width $\Gamma_t=1.33$ GeV; far above threshold, the ratio $K_R/\beta$ reproduces the Sommerfeld factor. The colour--singlet QCD background, the $A$ boson signal and the interference are all in the ${}^1S_0$ singlet channel and are multiplied by the same $K_1$,
\beq
\frac{d\sigma_1}{dm_{t\bar t}}\propto\left|{\cal Q}(\hat s)+\frac{{\cal N}_A(\hat s)}{\hat s-M_A^2+iM_A\Gamma_A(\hat s)}\right|^2K_1(E)\, ,
\eeq
with ${\cal Q}$ the short--distance QCD amplitude and ${\cal N}_A$ the $A$ residue: the toponium peak and the $A$ resonance multiply the same Green function, and the $A$ amplitude interferes with the bound--state region below threshold.

The $A\to t\bar t$ partial decay width is subject to the same Coulomb interaction. Unitarity requires $\Gamma_A(\hat s)={3y_t^2}/({8\pi}) \times m_{t\bar t}\,K_1(E)$ with $y_t=g_{Att}m_t/v$, and correspondingly ${\rm Im}\,F_A\to{\rm Im}\,F_A\,K_1/\beta$ for the absorptive part of the $gg\to A$ form factor; otherwise the branching ratio $A\to t\bar t$ would exceed one. This is the Coulomb part of the large QCD correction to $\Gamma_A$ noted in Ref.~\cite{DEQ:2024}. Since the width and the absorptive part are rescaled by the same factor, ${\rm Im}\,\kappa=-1$ remains exact, while $|{\rm Re}\,\kappa|$ decreases as the width increases. The widths increase by factors of 1.2--3.5 (as shown in Table~\ref{tab:bench}), and $|\kappa|\simeq1.5$--$1.7$ for the three masses (see Table~\ref{tab:kappa}): the $A$ is in the regime of maximal distortion, whatever its coupling.

The resummation is applied for $E<15$ GeV and smoothly matched to fixed-order calculations for $E>45$ GeV, the matching window being varied to $(10,30)$ and $(25,65)$ GeV. The Coulomb coupling is evaluated at $\mu_C={\rm max}(32~{\rm GeV},m_t\beta)$, giving $\alpha_s\simeq0.14$ near threshold, and varied by $\pm15\%$. We use the NNPDF2.3 LO parton densities~\cite{NNPDF} with $\mu_F=\mu_R=m_t$ and $m_t=172.5$ GeV at $\sqrt s=13$ TeV. All results are at LO for the hard process; the QCD $K$--factors, about 2 for the signal and 1.5 for the background, are not applied, so absolute cross sections should be understood as relative.\smallskip

\subsection*{3. Numerical results}

\begin{figure}[!h]
\centering
\includegraphics[width=\textwidth]{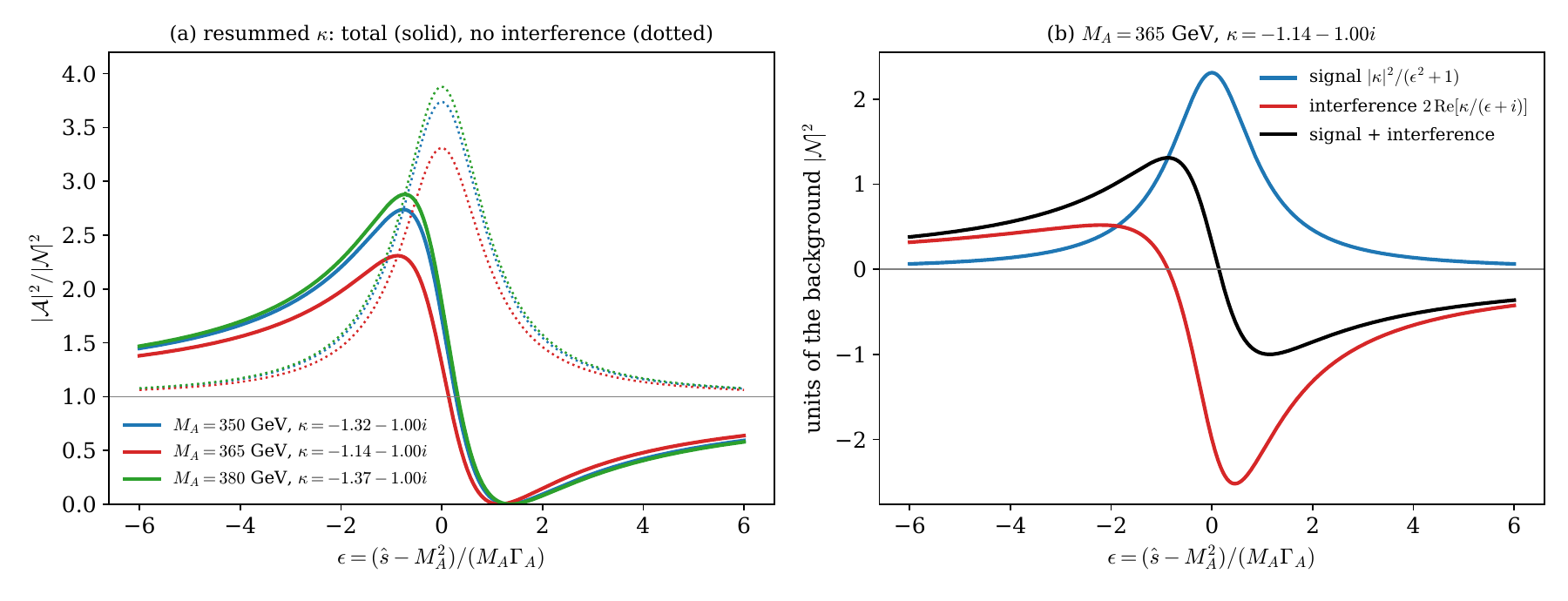}
\vspace*{-1cm}
\caption{(a) The Fano profile $|{\cal A}|^2/|{\cal N}|^2=|1+\kappa/(\epsilon+i)|^2$ in the colour--singlet ${}^1S_0$ channel for the resummed values of $\kappa$ of the three benchmarks (solid), compared with the incoherent sum of signal and background (dotted). (b) Decomposition for $M_A=365$ GeV, $\kappa=-1.14-1.00\,i$, in units of the background: signal, interference and their sum.}
\label{fig:fano}
\vspace*{-3mm}
\end{figure}

{We are now in a position to present our numerical results for the interference.} The line shapes for the resummed values of $\kappa$ are shown in Fig.~\ref{fig:fano}: since $\kappa$ does not depend on $g_{Att}$, they hold for all the benchmarks of Table~\ref{tab:bench}, the variable $\epsilon$ simply scaling with the width. For all three masses, the resonance is strongly distorted: the peak is shifted below $M_A$, towards the toponium region, it is followed by an exact zero of the ${}^1S_0$ rate at $\epsilon=-{\rm Re}\,\kappa\approx1.1$--$1.4$, and the total rate in this channel is reduced by 70--90\% compared with the signal alone. The right panel decomposes the profile for $M_A=365$ GeV: the antisymmetric interference term is positive below $M_A$ and negative above it, and its symmetric part removes most of the signal.

At fixed order and for the benchmarks with $f\!=\!100\%$, we reproduce the widths of Ref.~\cite{DEQ:2024} and, with the $K$--factors used there, the ratio of signal--only to signal--plus--interference rates in the narrow window $345$--$375$ GeV at $M_A\!=\!365$ GeV, 1.20. For the wide window $345$--$600$ GeV, we find $\approx 3$ instead of $\approx4$; this ratio is dominated by the far off--shell tail of the interference and depends strongly on its treatment: if the $gg\to A$ form factor is frozen at its on--shell value instead of evaluated at $m_{t\bar t}$, the ratio is $\approx 6$.

\begin{figure}[!h]
\centering
\includegraphics[width=0.48\textwidth]{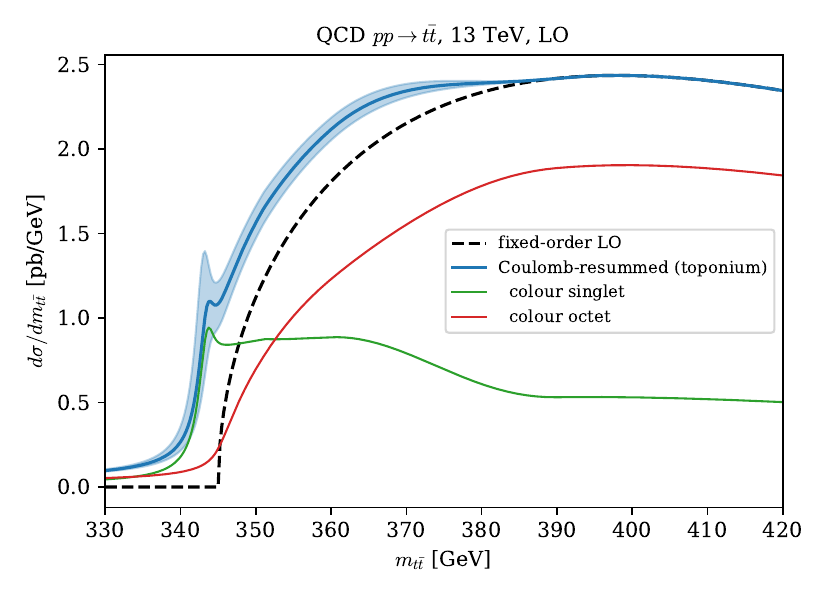}\hfill
\includegraphics[width=0.48\textwidth]{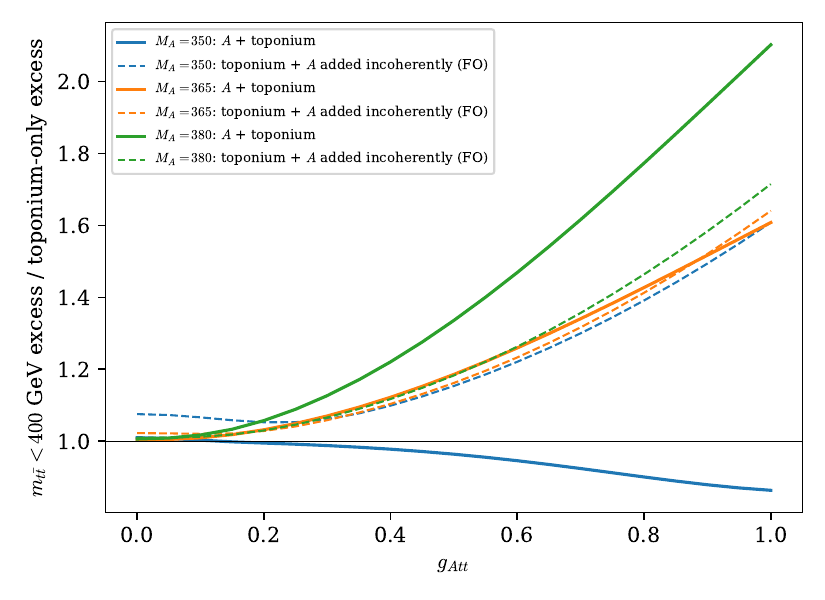}
\vspace*{-2mm}
\caption{Left: QCD $t\bar t$ production near threshold at fixed order and with Coulomb resummation (band: $\alpha_s^{\rm Coul}\pm15\%$), separated into colour singlet and octet. The right panel shows the threshold excess over fixed--order QCD ($m_{t\bar t}<400$ GeV) with an $A$ boson, normalized to the excess from toponium alone, as a function of $g_{Att}$. The solid lines are for  coherent calculations including the interference with toponium and the dashed lines are for toponium and $A$ added incoherently. The benchmarks with $f=100\%$, 50\% and 25\% correspond to $g_{Att}\approx0.8$, 0.6 and 0.4 (see Table~\ref{tab:bench}).}
\label{fig:qcd}
\end{figure}

The resummed QCD prediction, Fig.~\ref{fig:qcd} (left), shows the $\eta_t$ peak about 1.3 GeV below $2m_t$; the colour--singlet enhancement extends well above threshold, while the octet is suppressed. The excess over fixed-order calculations in the range $330$--$350$ GeV is about 8 pb (6--10 pb including the $\alpha_s$ variation) at LO, of the same size as the excesses observed by CMS and ATLAS~\cite{CMS:toponium,CMS:ljets,ATLAS:toponium} in the dilepton channel. In the single--lepton channel, CMS finds an enhancement of about 3 pb for $m_{t\bar t}<350$ GeV~\cite{CMS:ljets}.

\begin{figure}[!h]
\centering
\includegraphics[width=\textwidth]{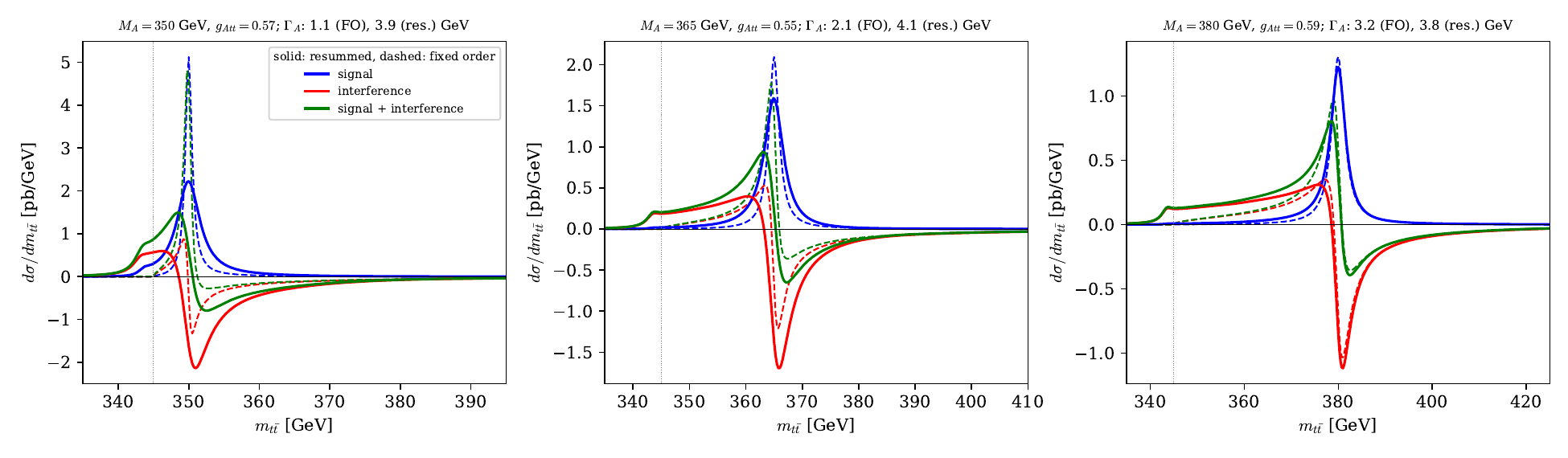}
\vspace*{-2mm}
\caption{Signal (blue), interference (red) and their sum (green) for the benchmarks with $f=50\%$ ($g_{Att}=0.57$, 0.55 and 0.59 for $M_A=350$, 365 and 380 GeV), at fixed order (dashed) and with Coulomb resummation (solid). The vertical line marks $2m_t$.}
\label{fig:SI}
\end{figure}

Figure~\ref{fig:SI} compares the fixed--order and resummed signal and interference for the benchmarks with $f=50\%$, and Table~\ref{tab:res} gives the integrated values for the reference benchmarks with $f=100\%$. The interference cross-section includes a new, positive contribution below threshold, absent at fixed order, where the off--shell $A$ amplitude is combined with the Coulomb--enhanced singlet continuum and the bound state. Below $m_{t\bar t}=347$ GeV it amounts to $+3.7$, $+2.1$ and $+1.4$ pb for $f=100\%$, compared with $+0.6$, $+0.1$ and $+0.1$ pb at fixed order, and it is still $+3.2$, $+1.1$ and $+0.7$ pb for $f=50\%$ (see Table~\ref{tab:frac}). Above the resonance, the destructive interference is enhanced as well. The integrated values in the windows of Table~\ref{tab:res} differ from the narrow--width $\delta$ of Table~\ref{tab:kappa}, since the windows are asymmetric around $M_A$ and include the positive lobe below threshold, where the spectrum falls steeply; this is the effect of the second term discussed in Ref.~\cite{DE:2026}.

\begin{table}[!h]
\renewcommand{\arraystretch}{1.2}
\centering\small\setlength{\tabcolsep}{4pt}
\begin{tabular}{l cc cc cc}
\toprule
 & \multicolumn{2}{c}{~~~$(350,0.81$)~~~} & \multicolumn{2}{c}{~~~($365,0.78$)~~~} & \multicolumn{2}{c}{~~~$(380,0.83$)~~~}\\
 & FO & res. &FO & res. &FO & res.\\
\midrule
$\Gamma_A$ [GeV] & 2.3 & 7.9 & 4.3 & 8.2 & 6.4 & 7.6\\
$S$ in $[330, M_A+10]$ [pb] & 16.2 & 20.4 & 12.3 & 16.6 & 11.3 & 13.7\\
$I$ in $[330, M_A+10]$ [pb] & $-4.7$ & $-13.9$ & $-2.6$ & $-6.1$ & $-1.5$ & $+1.1$\\
$I$ for $m_{t\bar t}<347$ GeV [pb] & $+0.6$ & $+3.7$ & $+0.1$ & $+2.1$ & $+0.1$ & $+1.4$\\
$(\!B\!+\!S\!+\!I)/B$ at $\eta_t$ peak & -- & 2.17 & -- & 1.38 & -- & 1.25\\
\bottomrule
\end{tabular}
\caption{Fixed--order (FO) and Coulomb--resummed results at LO (no $K$--factors) at $\sqrt s=13$ TeV, for the reference benchmarks with the couples $(M_A$ in GeV, $g_{Att})$ and $f=100\%$. $S$: signal; $I$: interference; $B$: QCD background.}
\label{tab:res}
\end{table}

\begin{figure}[!h]
\centering
\includegraphics[width=\textwidth]{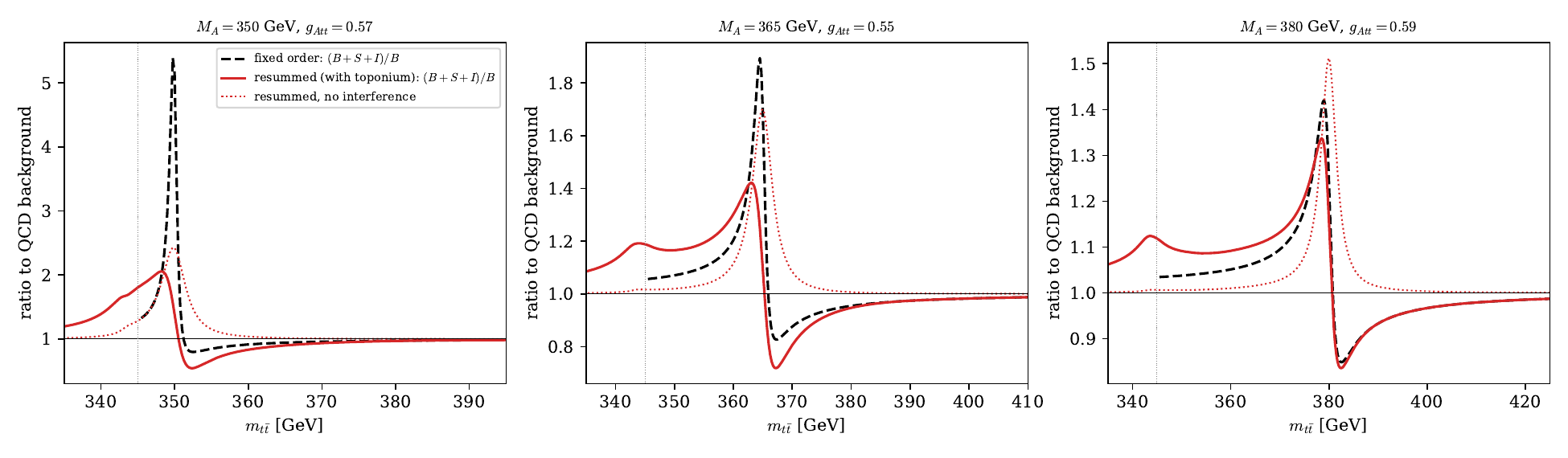}
\vspace*{-6mm}
\caption{Ratio of the total rate to the QCD background for the benchmarks with $f=50\%$, at fixed order (dashed) and with resummation (solid; dotted: without interference). With resummation, the $A$ enhances the toponium peak below $2m_t$ (vertical line).}
\label{fig:ratio}
\vspace*{4mm}
\end{figure}

Figure~\ref{fig:ratio} shows the ratio of the total rate to the QCD background for the benchmarks with $f=50\%$, and Table~\ref{tab:frac} gives the enhancement of the rate at the $\eta_t$ peak for the three fractions. With resummation, an $A$ boson accounting for half of the excess increases the rate at the $\eta_t$ peak by factors of approximately 1.7, 1.2 and 1.1 for $M_A=350$, 365 and 380 GeV; for $f=25\%$, the factors are still 1.4, 1.1 and 1.05, and they reach 2.2, 1.4 and 1.25 for $f=100\%$. The reason is that toponium lies below $M_A$, where ${\rm Re}\,\kappa<0$ makes the interference positive. These factors are robust: they vary by less than $\pm 20\%$ under the systematic variations of Table~\ref{tab:sys}.

\begin{table}[!h]
\renewcommand{\arraystretch}{1.25}
\centering\small\setlength{\tabcolsep}{4.5pt}
\begin{tabular}{l ccc ccc ccc}
\toprule
 & \multicolumn{3}{c}{$M_A=350$ GeV} & \multicolumn{3}{c}{$M_A=365$ GeV} & \multicolumn{3}{c}{$M_A=380$ GeV}\\
$f$ & 100\% & 50\% & 25\% & 100\% & 50\% & 25\% & 100\% & 50\% & 25\%\\
\midrule
$S$ in $[330,M_A+10]$ [pb] & 20.4 & 11.8 & 6.4 & 16.6 & 9.2 & 4.8 & 13.7 & 7.1 & 3.6\\
$I$ in $[330,M_A+10]$ [pb] & -13.9 & -8.3 & -4.5 & -6.1 & -3.5 & -1.9 & +1.1 & +1.0 & +0.6\\
$I$ for $m_{t\bar t}<347$ GeV [pb] & +3.7 & +3.2 & +2.0 & +2.1 & +1.1 & +0.6 & +1.4 & +0.7 & +0.4\\
$(B+S+I)/B$ at the $\eta_t$ peak & 2.17 & 1.69 & 1.36 & 1.38 & 1.19 & 1.10 & 1.25 & 1.12 & 1.06\\
excess/toponium, coherent & 0.90 & 0.95 & 0.98 & 1.41 & 1.22 & 1.12 & 1.82 & 1.45 & 1.24\\
excess/toponium, incoherent & 2.10 & 1.57 & 1.29 & 1.85 & 1.44 & 1.22 & 1.77 & 1.41 & 1.21\\
\bottomrule
\end{tabular}
\caption{Coulomb--resummed results at LO and $\sqrt s=13$ TeV for $A$ bosons accounting for the indicated fractions $f$ of the threshold excess (Table~\ref{tab:bench}): integrated signal and interference, enhancement of the rate at the $\eta_t$ peak, and threshold excess over fixed--order QCD ($m_{t\bar t}<400$ GeV) normalized to that of toponium alone, in the coherent calculation and when the fixed--order $A$ signal is added incoherently to toponium.}
\label{tab:frac}
\end{table}

Figure~\ref{fig:qcd} (right) shows the excess over fixed--order QCD for $m_{t\bar t}<400$ GeV, normalized to that from toponium alone, as a function of $g_{Att}$, and Table~\ref{tab:frac} compares the coherent calculation with the incoherent sum of toponium and of the fixed--order $A$ signal for the three fractions. The differences are large. For $M_A=350$ GeV, an $A$ boson accounting for 50\% of the excess would hardly change the threshold excess (factor 0.95), while the incoherent sum would increase it by 57\%: the signal is almost entirely cancelled by the destructive interference above $M_A$, which falls inside the threshold region. For $M_A=365$ GeV, the coherent result (1.22) is smaller than the incoherent sum (1.44), while for $M_A=380$ GeV the two are similar (1.45 and 1.41), the positive lobe below threshold compensating the destructive interference above $M_A$. The pattern is the same for $f=25\%$ and 100\%. Fits of the threshold excess with toponium and an $A$ boson should therefore include their interference, which can make an $A$ boson with a sizable coupling nearly invisible in the threshold region.

\begin{figure}[t]
\centering
\includegraphics[width=\textwidth]{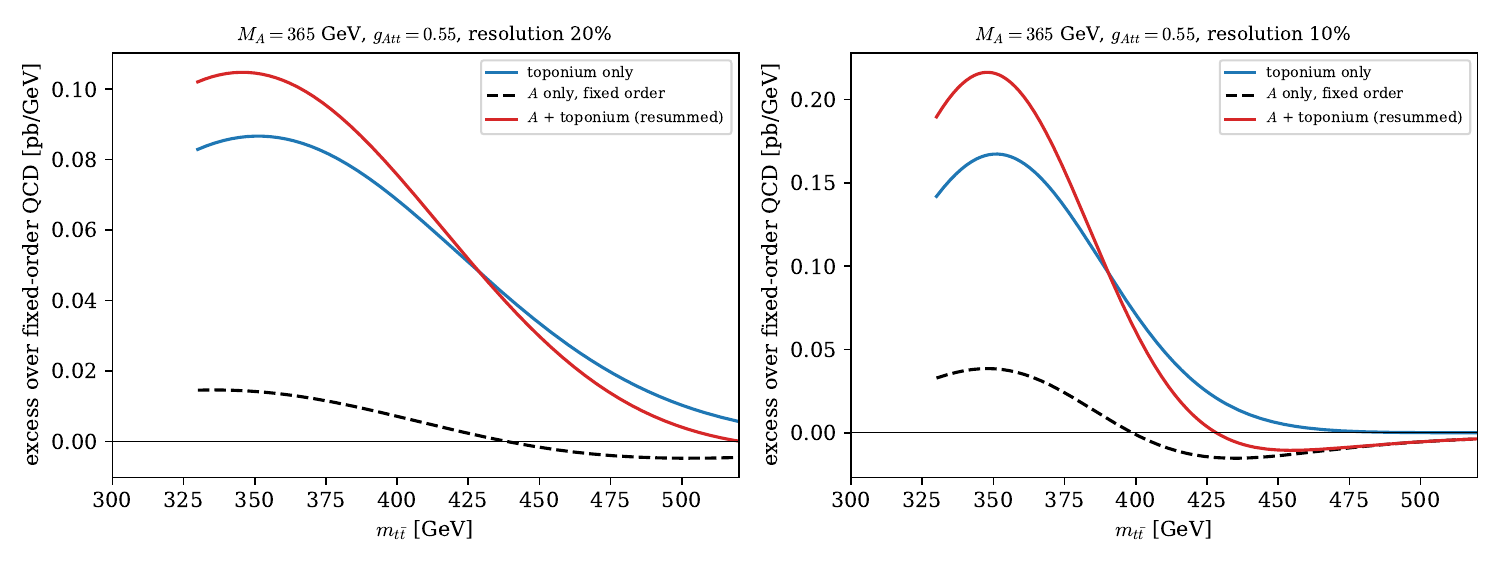}
\caption{Excess over fixed--order QCD after Gaussian smearing with 20\% (left) and 10\% (right) resolution, for $M_A=365$ GeV and $g_{Att}=0.55$ ($f=50\%$): toponium alone (blue), the $A$ alone at fixed order (dashed), and $A$ plus toponium (red).}
\label{fig:smeared}
\end{figure}

Figure~\ref{fig:smeared} shows the excess after Gaussian smearing, for $M_A=365$ GeV and $f=50\%$. With 20\% resolution, toponium alone and toponium plus $A$ have very similar shapes and differ mainly in normalization. With 10\%, the high--mass side of the $A$ resonance, where the interference is destructive, becomes visible as a faster fall--off. The fixed--order $A$ alone, as in our previous analysis, gives a smaller and differently shaped excess.

\begin{table}[!h]
\renewcommand{\arraystretch}{1.15}
\centering\small
\begin{tabular}{l c ccc ccc}
\toprule
 & toponium & \multicolumn{3}{c}{$\Gamma_A$ [GeV]} & \multicolumn{3}{c}{excess ratio, $m_{t\bar t}<400\,$ GeV}\\
 & $[330,350]$ [pb] & 350 & 365 & 380 & 350 & 365 & 380\\
\midrule
central & 7.9 & 7.9 & 8.2 & 7.6 & 0.90 & 1.41 & 1.82\\
matching $(10,30)$ GeV & 7.9 & 7.9 & 6.4 & 6.4 & 1.01 & 1.77 & 1.99\\
matching $(25,65)$ GeV & 7.9 & 7.9 & 8.5 & 10.4 & 0.84 & 1.22 & 1.49\\
$\alpha_s^{\rm Coul}\times0.85$ & 6.2 & 6.8 & 7.5 & 7.4 & 1.01 & 1.52 & 1.97\\
$\alpha_s^{\rm Coul}\times1.15$ & 10.0 & 8.9 & 8.9 & 7.8 & 0.81 & 1.32 & 1.71\\
\bottomrule
\end{tabular}
\caption{Systematic variations of the matching window and of the Coulomb coupling, for the benchmarks with $f=100\%$. The enhancement of the $\eta_t$ peak varies by less than 15\% (1.8--2.4, 1.32--1.38 and 1.19--1.25 for the three masses).}
\label{tab:sys}
\vspace*{-3mm}
\end{table}

\subsection*{4. Conclusions}

A pseudoscalar boson $A$ coupling to top quarks near the $t\bar t$ threshold necessarily interferes with toponium, because both appear in the colour--singlet ${}^1S_0$ channel and share the same Coulomb final--state interaction. Describing the line shape with the interference parameter $\kappa$, as in our analysis of the MSSM Higgs channels~\cite{DE:2026}, our findings are as follows. 

-- Unitarity fixes ${\rm Im}\,\kappa=-1$ exactly when $A$ decays into top quarks: the interference reduces the resonant rate by a factor $2/|\kappa|^2$, and the ${}^1S_0$ amplitude has an exact (Fano) zero just above the nominal $M_A$ value. All the model dependence resides in ${\rm Re}\,\kappa$.

-- The  width of the pseudoscalar $A$ boson is Coulomb--enhanced by factors of 1.2--3.5 near threshold, which brings $|\kappa|$ down to 1.5--1.7: the resonance is in the regime of maximal distortion, broader, lower and shifted below $M_A$.

-- $\kappa$ does not depend on the $A$ coupling to top quarks, so that these features hold for an $A$ boson accounting for any fraction of the observed excess.

-- The $A$ boson interferes constructively with the toponium peak, enhancing it by about 10--70\% (5--35\%) if it accounts for 50\% (25\%) of the excess, depending on $M_A$.

-- The threshold excess including $A$ is not the sum of the toponium and $A$ contributions. For $f=50\%$, it ranges from 0.95 to 1.45 times the toponium excess, compared with 1.4--1.6 for the incoherent sum. An $A$ boson with $M_A\approx350$ GeV can be nearly invisible. Fits of the threshold excess with an $A$ and the $\eta_t$ should therefore include their interference.

-- The wide--window ratio of signal to signal plus interference proposed in our previous analysis as a probe of the interference is sensitive to the off--shell treatment of $A$, while the narrow--window ratio is robust.\smallskip

The hard processes are calculated at LO, without $K$--factors, and the Coulomb potential is also taken at LO, with a constant top width; the matching to fixed order is a smooth interpolation, varied as a systematic; and top decays and spin correlations are not simulated. A next step would be to combine this framework with higher-order QCD calculations for both the signal and the background, see for instance the results in Refs.~\cite{HO-signal},
and with a full bound--state treatment including the spin correlations.

{The theoretical tools for searches for spin-0 resonances in $t \bar t$ final states at the LHC are well developed, and were used in the discovery by {the CMS Collaboration} of a pseudoscalar enhancement just below the $t \bar t$ threshold~\cite{CMS:HIG22013}. According to Occam's razor, this enhancement is most plausibly due to the pseudoscalar toponium state, $\eta_t$, rather than an elementary boson $A$. However, this cannot be regarded as established, and there is also the possibility that an $A$ boson could have a mass close to the $\eta_t$ state~\footnote{As the Freanch would say, ``un train peut en cacher un autre", one train can hide another.}. In this paper we have presented a formalism for analysing this possibility including the full interference effects, which are substantial. The $\eta_t$ peak may be enhanced significantly and the $A$ peak may be strongly distorted. We hope that our results will prove useful for future experimental studies of the region around the $t \bar t$ threshold to constrain these possibilities.} \bigskip

\noindent {\bf Acknowledgements}: 
We thank Claude Anthropic for valuable technical help. A.D. is supported by the Spanish grant PID2024-161668NB-I00 and thanks  CERN for hospitality.

\end{document}